\documentclass[twocolumn,11pt]{proc}
\usepackage[dvips]{graphicx}

\begin{document}

\title{Analysis of Non-Gaussian Nature of\\ Network Traffic}

\author{Tatsuya Mori\dag,\quad\quad Ryoichi Kawahara\ddag\quad\quad
Shozo Naito\dag\\
tatsuya@nttlabs.com, \{kawahara.ryoichi, naito.shozo\}@lab.ntt.co.jp\\
\dag NTT Information Sharing Platform Laboratories, NTT corporation\\
\ddag NTT Service Integration Laboratories, NTT corporation
}

\markboth{draft: IEICE General Conference 2002, Tokyo, Japan}
{draft: IEICE General Conference 2002, Tokyo, Japan}

\maketitle

\begin{abstract}
 To study mechanisms that cause the non-Gaussian nature of network traffic,
 we analyzed IP flow statistics.
 For greedy flows in particular, we investigated the hop counts between
 source and destination nodes, and classified applications by the port
 number.
 We found that the main flows contributing to the non-Gaussian nature
 of network traffic were HTTP flows with relatively small hop counts
 compared with the average hop counts of all flows.
\end{abstract}

\section{Introduction}
 Recently, it has been found that the characteristics of the marginal
 distribution of network traffic are crucial for modeling network traffic
 to evaluate performance\cite{mori-IN-01}.
 That is, marginal distributions are far from Gaussian and are skewed
 positively in many cases, and this nature is strongly correlated with
 network performance.
 It has also been found that this non-Gaussian nature of network traffic
 has a correlation with the heavy-tailedness of the per-time-block flow size
 distribution. 
 That is, according to the power-law of the distribution, some flows
 send a tremendously large number of packets in a given short time while
 most other flows send a rather small number of packets
 \cite{mori-IN-01}.
 As the nature of these greedy flows contributes to the non-Gaussian
 nature of network traffic, it is important to study their nature in
 detail.
 In this work, to investigate greedy flows, we studied hop
 counts and types of applications in them.
 
\section{Data}
 We used the trace data from the MAWI traffic archive\cite{mawi}
 measured at sample point-B between September and November 2001.
 The line is a 100-Mbps link with 18-Mbps CAR (committed access rate);
 it is one of the international lines of the WIDE project.
 All traces were measured during daily busy hours (14:00 - ) and
 contained about 2.9 $\sim$ 3.0 million packets.
 For this study, we used one-way US-to-Japan traffic because the average
 amount of traffic is much larger than in the opposite
 direction\footnote{To estimate hop counts, we
 needed to use traffic in both directions.}.
 For all traces, we calculated the average rate and skewness of traffic
 variability --- time series of throughput using the time interval of
 0.1 s.
 Total time average of variability for each trace varied from 6.02 Mbps
 to 34.70 Mbps (ensemble average was 18.80 Mbps).
 The skewness of variability for each trace varied from -0.61 to 2.71
 (ensemble average was 0.68).
 We removed traces with skewness smaller than 0.4 because our goal was to
 investigate the characteristics of network traffic having a non-Gaussian
 nature.
 In total, we used 68 traces for this work.

\section{Per-time-block flow analysis}
\begin{figure}[htbp]
 \begin{center}
  \includegraphics[width=80mm]{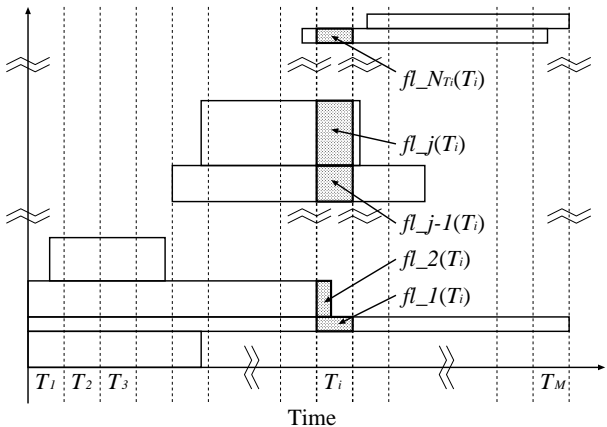}
  \caption{Diagram of per-time-block flow statistics.}
  \label{fig:flowstat}
 \end{center}
\end{figure}

\begin{figure}[htbp]
 \begin{center}
  \includegraphics[width=80mm]{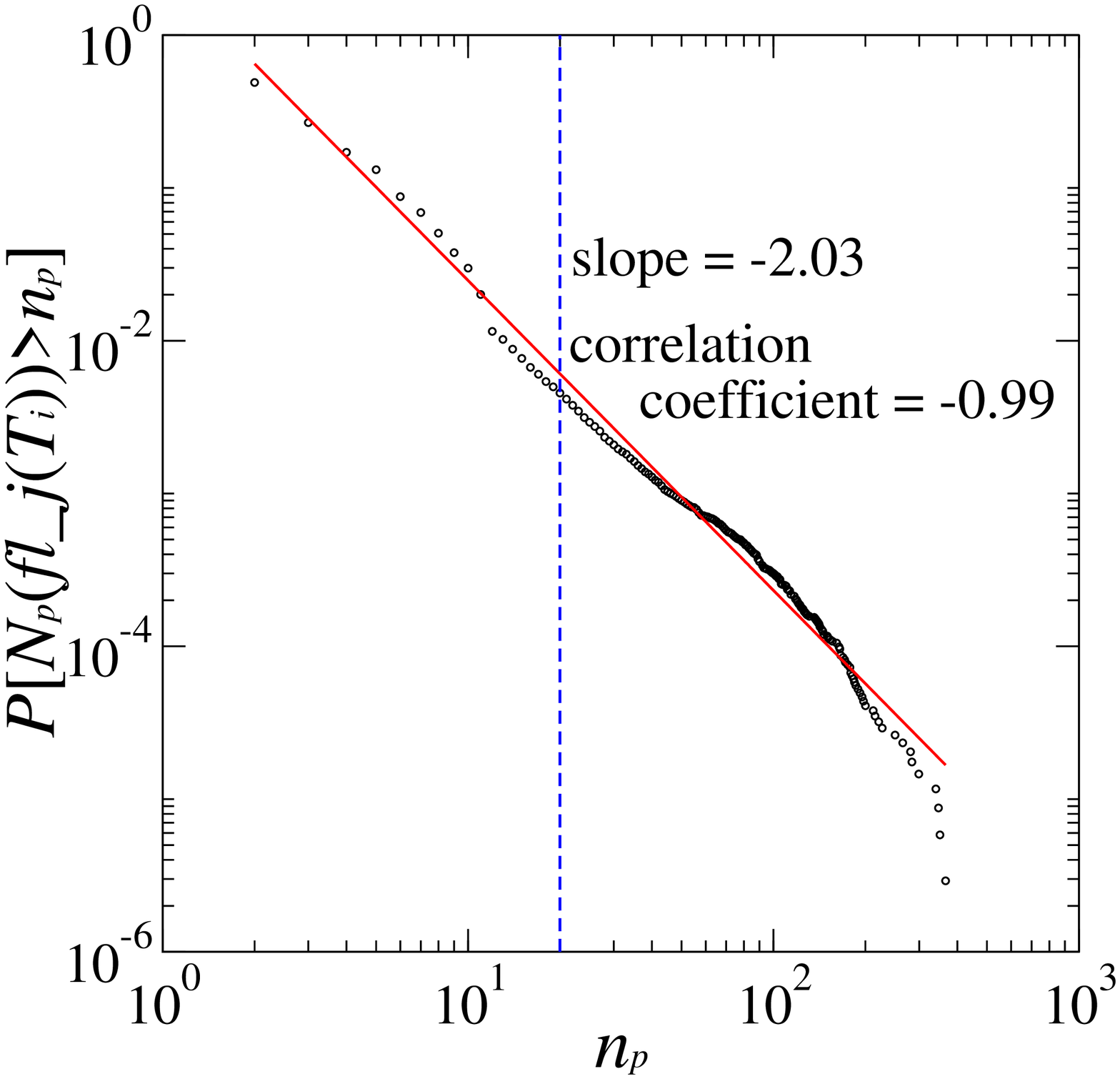}
  \caption{LLCD plots of $N_p\left(fl\_j\left(T_i\right)\right)$.}
  \label{fig:llcd}
 \end{center}
\end{figure}

We divided traces into time block $T_i$, where $1\leq i\leq M$ as
illustrated in Figure~\ref{fig:flowstat}. 
Here, for all $i$, the length of $T_i$ was set to time interval
$\tau$. For each $T_i$, we define a flow $fl\_j\left(T_i\right)$, where
$1\leq j \leq N_{T_i}$ and $N_{T_i}$ is the number of flows during
$T_i$.
Each flow $fl\_j\left(T_i\right)$ is defined as having an identical
combination of source IP address, destination IP address, source port,
destination port, and protocol.
The flow $fl\_j\left(T_i\right)$ should contain at least two packets
during $T_i$.
In this work, the length of $\tau$ was set to 0.1 s.
For each flow $fl\_j\left(T_i\right)$ $\left(1\leq j\leq
N_{T_i}\right)$, we counted the number of packets
$N_p\left(fl\_j\left(T_i\right)\right)$.
Figure~\ref{fig:llcd} shows the log-log complementary cumulative
distribution (LLCD) plots of $N_p\left(fl\_j\left(T_i\right)\right)$ for
all $i,j$.
These were in good agreement with the power-law as demonstrated in
\cite{mori-IN-01}.
In this work, we focused on greedy flows --- the tail part of the
distribution.
Here, we define a greedy flow as one whose
$N_p\left(fl\_j\left(T_i\right)\right)$ is larger than 20 (right side of
the dashed line in Figure~\ref{fig:llcd}), which corresponds to
throughput about 1 Mbps assuming the average packet size to be 700
bytes.

\section{Hop count estimation}
 \subsection{Estimation technique}
 To study hop counts between two nodes from the given trace data, we
 used the TTL (time to live) field of an IP packet. As its value is
 decreased when an IP packet passes a router, we can estimate hop counts
 between the source node and measuring point from the initial TTL value
 and the TTL value of the recieved packet.
 So, if we can obtain the hop counts from both the source and
 destination nodes to the measuring point, we can estimate the hop
 counts
 between these nodes
 \footnote{Here, we assumed that the routing paths for both
 directions were the same for convenience.}.
 One difficulty with this approach is that the initial TTL values
 depend on the operating system or network equipment such as
 routers (see \cite{lance99} for example).
 So first of all, we must determine the initial TTL value of source
 nodes. 
 In this work, we used the approach of passive OS fingerprinting
 to estimate hop counts as exactly as possible.
 This technique is based on the principle that every system has its own
 IP stack implementation.
 That is, we can detect systems extremely accurately using some values
 recorded in IP packets they sent. 
 More detailed information about passive OS fingerprinting can be found
 in \cite{lance99}.
 In this work, we modified the source code of p.0.f.\cite{p0f} and
 estimated the hop counts of each IP flow.
 In our study, we could estimate more than 10\% of the systems
 for each trace.
 Here, we assume that we can regard statistics of these estimated
 flows as statistics of all flows.

 \subsection{Hop counts of greedy flows}
\begin{figure}[htbp]
 \begin{center}
  \includegraphics[width=80mm]{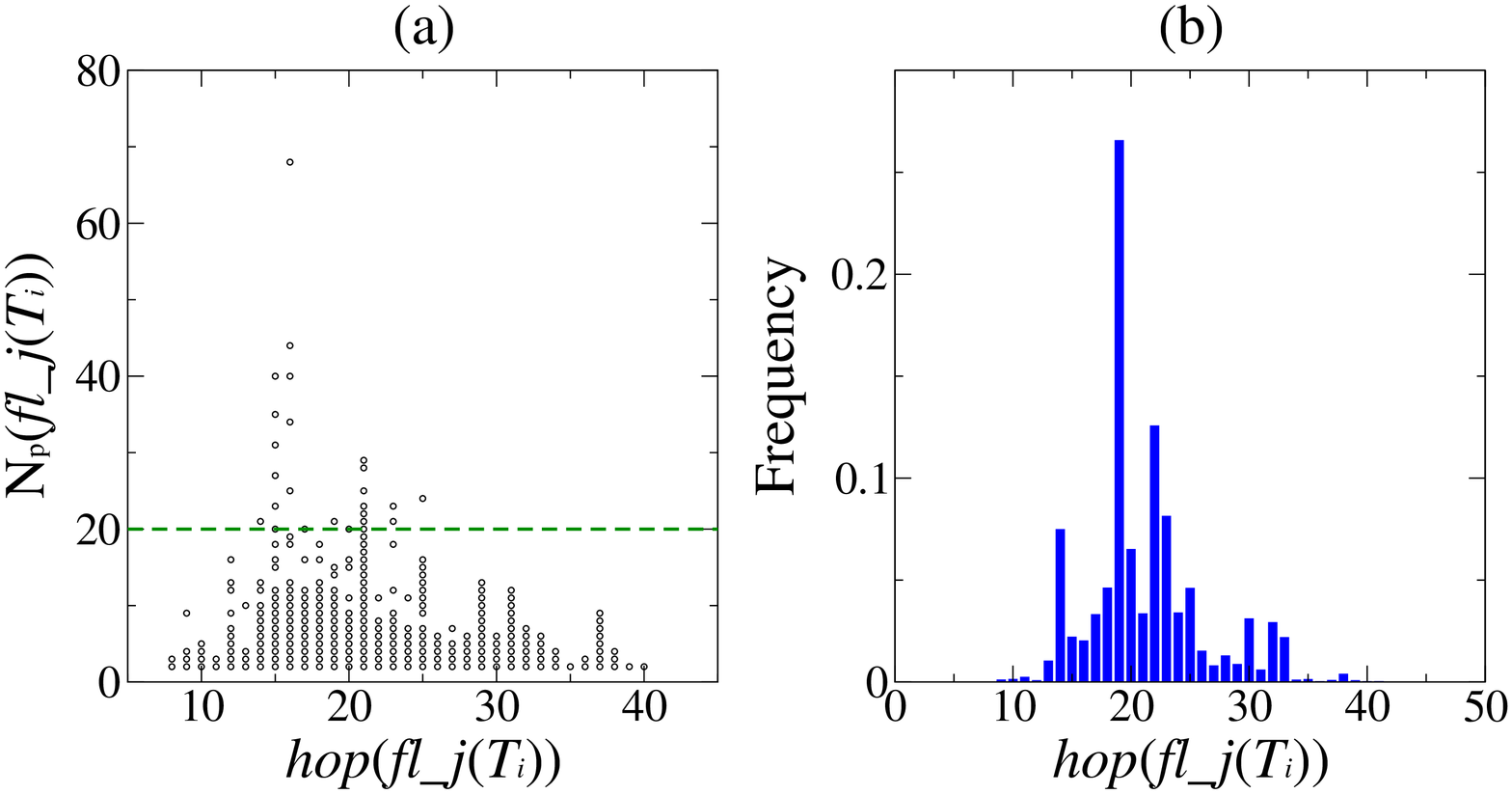}
  \caption{(a)  $hop\left(fl\_j\left(T_i\right)\right)$ versus
  $N_p\left(fl\_j\left(T_i\right)\right)$, (b) histogram of $hop\left(fl\_j\left(T_i\right)\right)$.}
  \label{fig:hop-numpkt-pdf}
 \end{center}
\end{figure}

 For each flow $fl\_j\left(T_i\right)$ defined in
 Sec. 3, we estimated hop counts $hop\left(fl\_j\left(T_i\right)\right)$
 using the above technique.
 Figure~\ref{fig:hop-numpkt-pdf}(a) shows the relationship between
 $hop\left(fl\_j\left(T_i\right)\right)$ and
 $N_p\left(fl\_j\left(T_i\right)\right)$ for a certain trace.
 Figure~\ref{fig:hop-numpkt-pdf}(b) shows the histogram of
 $hop\left(fl\_j\left(T_i\right)\right)$ for the same trace.
 The hop counts of greedy flows (above the dashed line in (a)) can 
 be considered to be smaller than those of all flows.

\begin{figure}[htbp]
 \begin{center}
  \includegraphics[width=80mm]{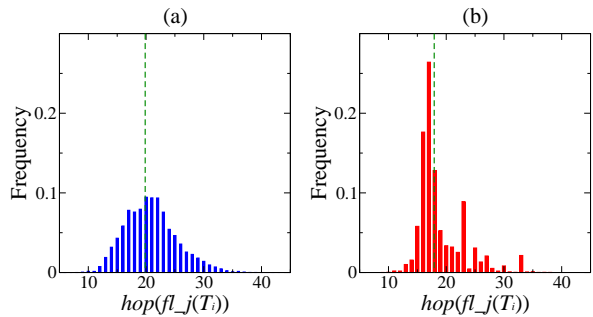}
  \caption{Histogram of $hop\left(fl\_j\left(T_i\right)\right)$
  for (a) all flows, (b) greedy flows.}
  \label{fig:all-total-greedy-hop-pdf}
 \end{center}
\end{figure}

 Then we investigated the histogram of
 $hop\left(fl\_j\left(T_i\right)\right)$ for all traces.
 Figure~\ref{fig:all-total-greedy-hop-pdf} shows histograms of (a) all
 flows  and (b) greedy flows.
 The average hop counts for greedy flows were smaller than those of all
 flows, and most greedy flows had relatively smaller hop counts. 
 Actually, average hop counts were 19.85 for all flows and 17.92 for
 greedy flows (see dashed lines in Figure~\ref{fig:all-total-greedy-hop-pdf}).
 These results can be interpreted using the fact that the RTTs of flows with
 smaller hop counts tend to be smaller as demonstrated in
 \cite{fujii00}, and TCP flows with smaller RTTs can make their window
 sizes larger following the mechanism of TCP flow control.
 That is, TCP flows with smaller hop counts can make their window sizes
 larger, and be greedier.

 \section{Breakdown of applications}
 We investigated the breakdown of applications for each flow using the port
 numbers (Table~\ref{tab:app-proportion}).
 We can immediately see that the proportion of HTTP was
 much larger among greedy flows than all flows.
 So it might be reasonable to assume that HTTP plays an important role
 in making greedy flows.
 We will focus on causal mechanisms of why HTTP flows are greedier than
 other applications, and the relationship with hop counts in our next
 work.

\begin{table}[htbp]
 \begin{center}
  \caption[]{Breakdown of applications}
  \label{tab:app-proportion}
  \begin{tabular}{ccc}
   \hline
   & All flows & Greedy flows\\
   \hline
   HTTP & 54\% & 70\% \\
   other TCP & 38\% &20\% \\
   UDP & 7\% & 6\%\\
   other & 1\% & 4\%\\
   \hline
  \end{tabular}
 \end{center}
\end{table}

 \section{Summary}
 To study mechanisms that cause the non-Gaussian nature of network
 traffic, we investigated the properties of greedy IP flows.
 Our main findings are as follows.
 (1) Hop counts of greedy flows were relatively smaller than those of all
 flows.
 (2) HTTP was the main application for greedy flows.
 Since the most popular application on the Internet today is
 server-client type file-transfer applications such as WWW, we believe
 that the results of this work suggest the ubiquitous
 existence of greedy flows, which causes the non-Gaussian nature of network
 traffic.

\end{document}